\documentclass{amsart}
\pdfoutput=1

\usepackage[utf8]{inputenc}
\usepackage[T1]{fontenc}
\usepackage{lmodern}

\usepackage{amssymb}
\usepackage{mathtools}
\usepackage{amscd}
\usepackage[mathscr]{eucal}
\usepackage{enumitem}
\usepackage[hmargin=1in,vmargin=1in]{geometry}

\usepackage{graphicx}

\usepackage[roman]{sublabel}

\usepackage[labelfont=bf,font=small]{caption}
\usepackage{subcaption}

\usepackage{cite}
\usepackage[foot]{amsaddr}

\usepackage{verbatim}
\usepackage[usenames, dvipsnames]{color} 
\usepackage[]{algorithm2e}

\usepackage{siunitx}

\definecolor{cite_blue}{rgb}{0,0,0.97}
\definecolor{link_red}{rgb}{0.7,0,0}
\usepackage[colorlinks=true,citecolor=cite_blue,linkcolor=link_red]{hyperref}

\theoremstyle{plain}

\theoremstyle{definition}

\begin{document}		
\title{
Modeling mechanical interactions in growing populations of rod-shaped bacteria
}

\author{James J. Winkle$^{1}$}
\author{Oleg Igoshin$^{2,3}$}
\author{Matthew R. Bennett$^{2,4}$}
\author{Kre\v{s}imir Josi\'{c}$^{1,4,5,*}$}
\author{William Ott$^{1,*}$}

\address{Department of Mathematics, University of Houston$^{1}$}
\address{Department of Bioengineering, Rice University$^{2}$}
\address{Center for Theoretical Biological Physics, Rice University$^{3}$}
\address{Department of Biosciences, Rice University$^{4}$}
\address{Department of Biology and Biochemistry, University of Houston$^{5}$}

\thanks{$*$ = Corresponding author.}

\date{\today}

\maketitle
\thispagestyle{empty}

\begin{abstract}    
Advances in synthetic biology allow us to engineer bacterial collectives with pre-specified characteristics.
However, the behavior of these collectives is difficult to understand, as cellular growth and division as well as extra-cellular fluid flow lead to complex, changing  arrangements of cells within the population.
To rationally engineer and control the behavior of cell collectives we need theoretical and computational tools to understand their emergent spatiotemporal dynamics.
Here, we present an agent-based  model that allows growing cells to detect and respond to mechanical interactions.
Crucially, our  model couples the dynamics of cell growth to the  cell's environment: Mechanical constraints can affect cellular growth rate and a cell may alter its behavior in response to these  constraints.
This coupling  links the mechanical forces that influence cell growth and emergent behaviors in cell assemblies.
We illustrate our approach by showing how mechanical interactions can impact the dynamics of bacterial collectives growing in microfluidic traps.

\end{abstract}

\section{Introduction}	
To realize the full potential of synthetic biology, we need to be able to design assemblies of \emph{interacting} cells and organisms.  
Cooperating cells can specialize and assume different responsibilities within a collective~\cite{Wintermute2010}.  
This allows such bacterial consortia to outperform monocultures, both in terms of efficiency and range of functionality, as the collective can  perform computations and make decisions that are far more sophisticated than those of a single bacterium~\cite{Regot2011}. 

Recent advances in synthetic biology allow us to design multiple, interacting  bacterial strains, and observe them over many generations~\cite{Chen2015}.  
However, the dynamics of such microbial consortia are strongly affected by spatial and temporal changes in the densities of the interacting strains. 
The spatial distribution of each strain determines the concentration of the corresponding intercellular signals across the microfluidic chamber, and in turn, the coupling among strains. 
To effectively design and control such consortia, it is necessary to understand the mechanisms that govern the spatiotemporal dynamics of bacterial collectives. 

Agent-based modeling provides an attractive approach to uncovering these mechanisms. Such models can capture behaviors and interactions at the single-cell level, while remaining computationally tractable.
The cost and time required for experiments make it difficult to explore the impact of inhomogeneous population distributions and gene activity under a variety of conditions.  
Agent-based models are far easier to run and modify. They thus provide a powerful method to generate and test hypotheses about gene circuits and bacterial consortia that can lead to novel designs. 

Importantly, agent-based models of microbial collectives growing in confined environments, such as microfluidic traps, should capture the effect of mechanical interactions between cells in the population.
Forces acting on the constituent cells play a critical role in the complex dynamics of cellular growth and emergent collective behavior~\cite{Drasdo2007, Shraiman2005, Si2015, Sun2011, Davidson2011}.
Agent-based models, therefore, need to be able to model the force exerted by growing cells, as well as the mechanical interactions induced by cell-cell contacts or contact with environmental boundaries.
Further, it has been shown that the  environment of an individual cell can influence its growth, which in turn influences the collective's behavior trough mechanical communication~\cite{Cho2007, Delarue2016, Fernandez-Sanchez2015, Sadati2013, Szab2006}.
In particular, mechanical confinement can cause cells within the collective to grow at different rates~\cite{Cho2007,Delarue2016}.
Current agent-based models of microbial collectives (e.g.~\cite{Hoehme2010 , Jang2012, Rudge2012, Gutierrez2017}) typically do not allow cells to alter their growth rate in direct response to mechanical sensory input.
Adding such capability is challenging, due to the complex relationship between cell growth and the extracellular environment.

Here, we introduce an agent-based bacterial cell model that can detect and respond to its mechanical environment.
We show that our model can be used to make predictions about the spatiotemporal dynamics of consortia growing in two-dimensional microfluidic traps.
Further, we demonstrate that emergent collective behavior can depend on how individual cells respond to mechanical interactions.

\begin{figure} 
\centering
	\includegraphics[scale=0.59]{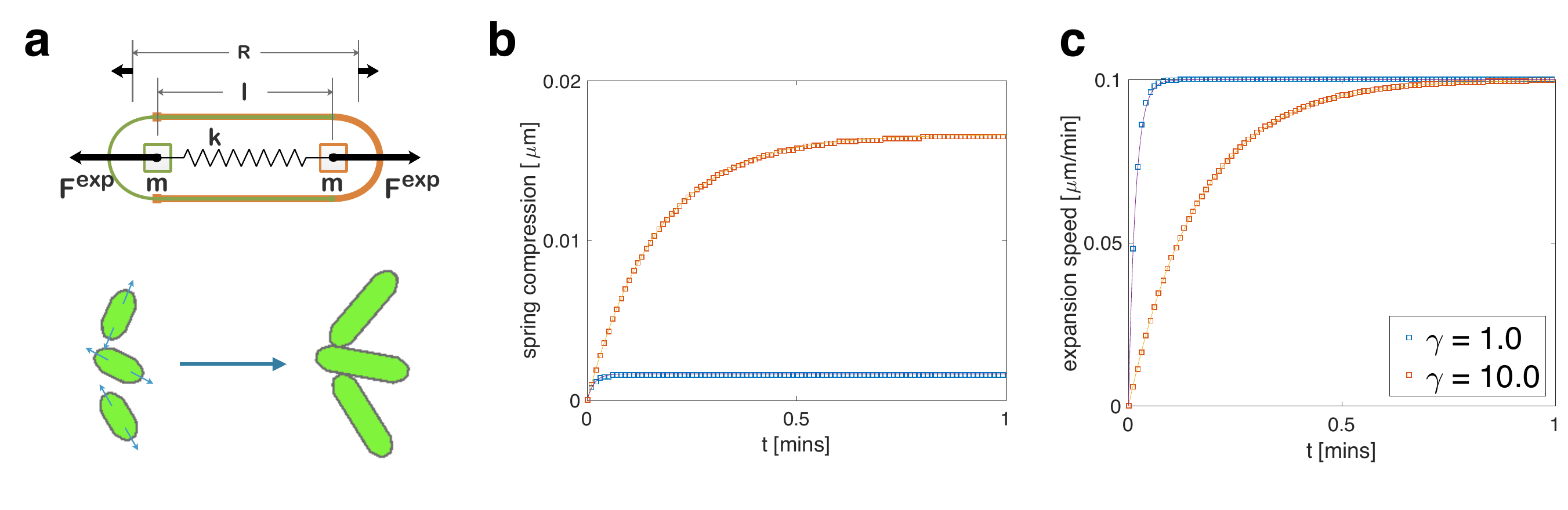}
\caption{\textbf{Single-cell construction and dynamics.}
\textbf{(a)} 
A schematic depiction of the components of a single cell in our model.
Two symmetric cell halves with semicircular poles and long edges are constrained to align using a ball-and-groove type connection. 
Each cell half has mass $m$ (assumed to remain constant during cell growth) and center of mass located at the center of its semicircular pole.  
Growth expansion forces are generated by connecting the two cell halves to a virtual linear spring (with spring constant $k$) along the cell's long axis, and then extending the rest length, $R,$ of the virtual spring.
\textbf{(bc)}
Expansion speed (magnitude of the difference between the velocities of the two cell halves) and spring compression (difference between rest length $R$ and cell length $\ell$) for a single cell with $\dot{R} = \SI{0.1}{\micro\meter\per\minute}$ and a 10-fold increase in resistive damping parameter $\gamma$.  
Simulation data (boxes) match analytical solutions (curves).  
Increasing $\gamma$ (orange traces) lengthens the time required for expansion speed and spring compression to reach steady state ($\tau = \frac{\gamma}{2k}$ is the time constant for the first-order system). 
Larger steady-state spring compression monitors the increased mechanical load felt by the growing cell when the damping coefficient is higher.
}
\label{fig:IsolatedCell}
\end{figure}

\section{Modeling Framework}
\label{s:Framework}
To understand the behavior of growing bacterial collectives, we must develop numerical tools that can capture the  mechanisms that shape their spatiotemporal dynamics.
Here, we propose an agent-based model of bacterial assemblies, using a framework that takes into account mechanical constraints that can impact cell growth and influence other aspects of cell behavior. 
Taking these constraints into account is essential for an understanding of colony formation, cell distribution and signaling, and other emergent behaviors in cell assemblies growing in confined or crowded environments.

 Our framework differs from other published models in an important way: We assume that each cell comprises \emph{two axially independent cell halves} that attach through a compressible, stiff spring, whose rest length increases to induce cell growth (Fig.~\ref{fig:IsolatedCell}(a)). 
Our spring model serves as a first-order approximation of peptidoglycan cell wall response to mechanical stresses~\cite{Huang:2008}.  
The expansion rate of spring rest length sets the target growth rate for the cell.
However, in our model the growth rate may not be immediately achieved due to mechanical constraints, such as resistive damping, cell-cell contact, and contact with trap boundaries.
Differences in rest length expansion and actual cell growth result in sustained spring compressions, whose energy can be thought of as a  {\itshape stored growth potential} for the cell.


Most published models require that cells grow exactly  at  {\itshape a priori}  prescribed  rates.  An exception is a model introduced to study the organization of crowded bacterial colonies in growing in confined niches~\cite{Cho2007}.
As a result,  most models do not capture mechanical constraint detection and resultant growth modulation.
Our approach introduces greater flexibility than, for example, assuming that growth rate is determined by the position of a cell in a trap~\cite{Boyer2011}.

We first present our cell model and assumptions, derive the theoretical equations of motion for growth of an isolated cell, and validate simulation results by comparing them to our theoretical model.
In constructing our simulations and diagrams, we took advantage of two open-source software resources: the physics engine \texttt{Chipmunk 2D}~\cite{cp} for cell dynamics, and the cell simulation platform \texttt{gro}~\cite{Jang2012} (which we modified for use with our cell model) for visualizations and image sequencing.

\subsection					{Cell construction}
We model each bacterium as an assembly of two independent cell halves.  To model cell growth, we assume that these two halves expand symmetrically along the long axis of the bacterium (Fig.~\ref{fig:IsolatedCell}(a)).
Each cell half consists of a mass $m$ at the center of a semi-circular pole, which connects to straight, long-body edges (as shown by different colors in Figure~\ref{fig:IsolatedCell}(a)).  
The two masses connect through a virtual spring with linear spring constant $k$.  Importantly, the rest length of the spring increases in time.  
In confined environments, extension of the rest length induces forces on neighboring cells, microfluidic trap boundaries, and any other obstacles the cell may encounter.

In order to ensure the cell halves act as a single, well-defined cellular unit (for example, upon collision with other cells or fixed barriers), we use a pair of symmetric ball-and-groove type connections to ensure that the halves remain aligned and resist bending~\cite{Huang:2008}.  
This also ensures that any off-axis or rotational impulses are transmitted equally to both halves of the cell. 
Thus, cell growth forces are designed to act independently in the axial direction, whereas off-axis, cell-external forces act on the cell as a whole. 
Sufficiently large \emph{on-axis} components of external forces could result in a cell-length compression in this model, but we mediate this using a rigid-body, back-filling ``ratchet'' algorithm.  
Details about the implementation are  provided in the Appendix.

\subsection					{Growth model}

We induce axial cellular growth by extending the rest length, $R$, of the virtual spring that connects the cell halves (Fig.~\ref{fig:IsolatedCell}(a), top panel).
Induced expansion force can be felt by all neighboring objects (see Fig.~\ref{fig:IsolatedCell}(a), bottom panel).
Cho et al.~\cite[Figure~4]{Cho2007}  used a related model to study how mechanical constraints lead to self-organization 
in bacterial colonies grown in confined environments.

Crucially, rest length extension is an adjustable component of our model that captures the growth \emph{tendency} of each cell.
As we will see, altering how rest-length extension dynamics respond to constraint can impact global dynamics of collectives.
To start, however, in Sections~\ref{s:Framework} and~\ref{s:1Dmm} we assume that the rest length grows at a constant rate, $\dot{R} = a$.  
In this case, mechanical constraints can  result in unphysiologically large potential energy stored in a highly compressed spring, an issue we address in subsequent sections.

We assume cells grow in an extracellular fluid with a resistive damping parameter, $\gamma$, and that our system is in the non-inertial dynamics regime (see Appendix). 
Fluid damping resists cell growth via a damping force $\gamma \dot{x}$, where $\dot{x}$ is the lab-frame speed of a cell half through the extracellular fluid.
We explicitly model this parameter to explore the effects that fluid damping variations have on cell dynamics.  
Although $\gamma$ defines non-inertial dynamics over a broad range of values, we will see that it directly governs response dynamics under the assumptions of our growth model.
We make the simplifying assumption that $\gamma$ captures all sources of resistive damping, including extracellular fluid damping and dissipative (non-Hamiltonian) damping forces within the cell itself.
In particular, $\gamma$ serves as an imperfect but computationally manageable proxy for cell-internal spring damping.

Many published agent-based models treat bacterial cells as unitary rigid bodies under non-inertial dynamics that achieve cell growth by a process we will call the \emph{Expansion, Overlap, Relaxation} (EOR) method.
In these models, forward Euler integration of the growth rate $a$ expands (E) a cell by increasing its length by $a \cdot \mathrm{d}t$, where $\mathrm{d}t$ is the time discretization step. 
If a cell is sufficiently near, or in contact with, another object (for example another cell or a trap wall) just before this time step, expansion will result in overlap (O). 
A relaxation algorithm (R) is then asserted that resolves (or prevents) overlaps of all cells and objects using repulsion forces~\cite{Cho2007}, constraint~\cite{Rudge2012}, iteration~\cite{Jang2012}, or a related algorithm~\cite{Gutierrez2017}.

In our model, we prevent cell overlap by using \emph{collision dynamics} to resolve competing growth expansion under the constraint of cell-cell or cell-barrier contact (see Appendix).
Importantly, by constructing a cell with two axially independent halves, we do not have to assume that each cell reaches a predetermined size, determined by the growth rate, at the end of each time step.
In contrast to the EOR method, this allows us to determine the impact of mechanical constraints on the growth of a cell by comparing the achieved cell length $\ell$ to spring rest length $R$ at each time step.  
We can then link this measurement (which is made locally by the cell agents themselves), to other aspects of the cell model.
As we will see,  emergent assembly behavior can depend on how cells modulate growth in response to  constraints.

\subsection			{Equations of motion for an isolated cell}
\label{ss:eom}
We derive the equations of motion for an isolated cell in an extracellular fluid with resistive damping parameter $\gamma$.  
Cell growth results from a linear spring force computed from the difference between $\ell$ and the rest length $R$ of our virtual spring ($R - \ell$ is thus spring \emph{compression}) and applied to each cell half.
Using linear spring constant $k$, we have the inertial equation of motion for an isolated cell,
\begin{equation} \label{eqn:inertial}
	\frac{m \ddot{\ell}}{2} = k (R - \ell) - \frac{\gamma \dot{\ell}}{2}.
\end{equation}
Assuming non-inertial dynamics (see Appendix), Eq.~\eqref{eqn:inertial} yields a differential equation for expansion velocity,
\begin{equation}
\label{eqn:noninertial}
	\dot{\ell} = \frac{2k}{\gamma}(R - \ell).
\end{equation}
In order to close Eq.~\eqref{eqn:noninertial}, we must describe the dynamics of the rest length, $R$.
Bacteria grow approximately exponentially (see~\cite{Amir2014} and references contained therein).
However, for simplicity we let $R$ extend linearly at rate $a$, independent of cell length.  
This assumption can be relaxed, and does not affect the main points below.
In Section~\ref{s:2D}, we will introduce \emph{mechanical feedback} by modulating $\dot{R}$ in response to mechanical constraint.

Setting $R(0) = 0$, we have $R(t) = at$, so Eq.~\eqref{eqn:noninertial} becomes
\begin{equation} \label{eqn:ldotl}
	 \dot{\ell} + \frac{2k}{\gamma} \ell =  \frac{2k}{\gamma} a t.
\end{equation}
Defining $\tau = \frac{\gamma}{2k}$, setting initial cell length, $\ell,$ to zero,  and solving Eq.~\eqref{eqn:ldotl} gives  the length of an isolated cell, and the rate of its expansion,
\begin{equation} \label{eqn:ell}
	  \ell (t) = a ( t -\tau + \tau e^{- \frac{t}{\tau}}), \qquad 	  \dot{\ell} (t) = a (1 -  e^{- \frac{t}{\tau}}).
\end{equation}

The parameter $\tau$ acts as a time constant for growth dynamics.
Eq.~\eqref{eqn:ell} shows that $\dot{\ell} \to a$, and that $\tau$ governs the time required to reach steady state.
Since $\tau$ is proportional to resistive damping $\gamma$ for fixed $k$, resistive damping therefore governs this lag.
Using Eq.~(\ref{eqn:ell}), the compression of the spring that drives the growth of the isolated cell is given by
\begin{equation} \label{eqn:c}
	  R(t) - \ell (t) 
	  = at - a ( t -\tau + \tau e^{- \frac{t}{\tau}})
	  = a \tau (1 - e^{- \frac{t}{\tau}}).
\end{equation}
Notice that Eq.~(\ref{eqn:c}) implies that $(R - \ell) \to a \tau$, 
a measure of the sustained mechanical constraint felt by an isolated growing cell at steady state due to resistive damping.

As described in the Appendix, we have implemented this model using the \texttt{Chipmunk 2D} environment.
To validate our implementation, we first compared the growth of an isolated cell  to that given by Eq.~\eqref{eqn:ell}.
We varied resistive damping $\gamma$ by an order of magnitude, while using units such that $k=1$, and $\gamma$ was changed from $1$ to $10$.
Figure~\ref{fig:IsolatedCell}(bc) shows close agreement between theory and simulation for  spring compression and expansion speed.
The timescale at which both approach their equilibrium values increases with $\gamma$.

\section{Behavior of cells in a Mother Machine}
\label{s:1Dmm}

To bridge the divide between a single, isolated cell and collectives growing in general two-dimensional geometries, we now study a one-dimensional `mother machine' configuration, where cells are constrained to grow in long, narrow traps. 
Mother machines are microfluidic devices developed to study bacterial cell growth and division over hundreds of generations (See~\cite{Jun:MM,Wang2010}).   
They consist of an array of impermeable, three-walled narrow channels, each just wide enough to hold a line of cells. 
The open end of each channel is perpendicular to a `trench' through which fresh nutrient medium flows.
Cells exiting the narrow channels are carried away by this flow.

\begin{figure} 
\centering
				\includegraphics[scale=.61]{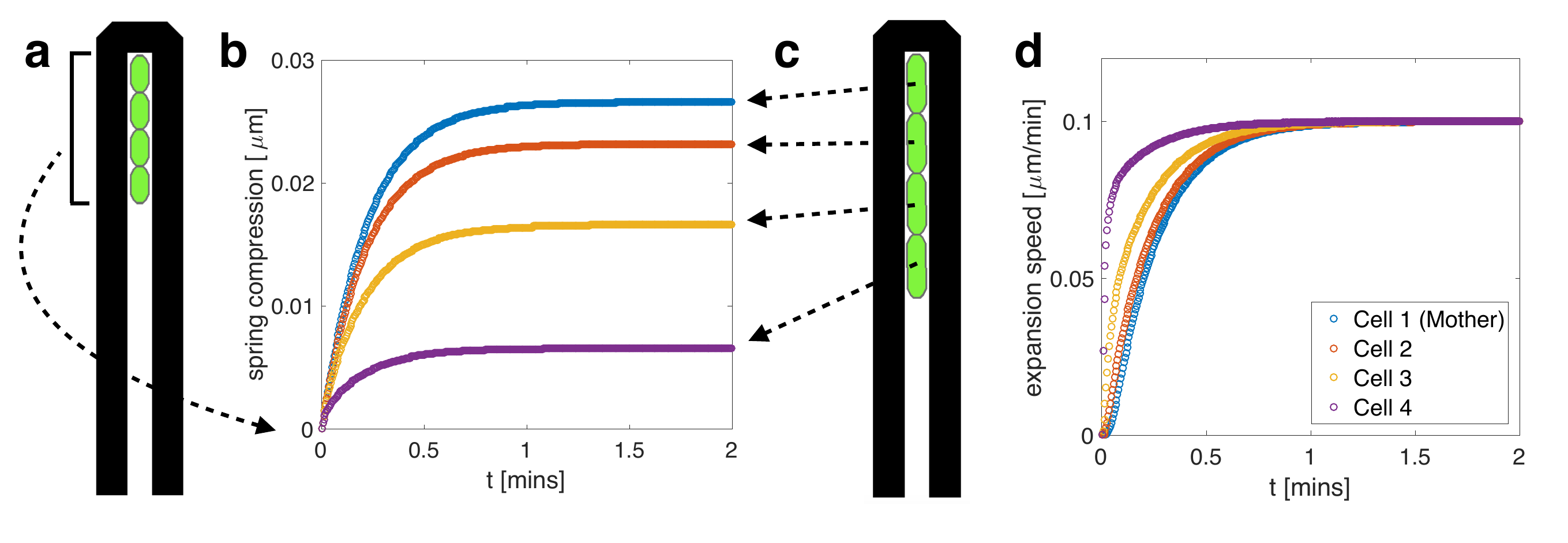}
\caption{
\textbf{1D mother machine trap simulation.} 
\textbf{(a)} 
A schematic depiction of the mother machine trap setup.
Four cells were placed back-to-back from the closed top of the trap and grew toward the open end~\textbf{(c)}.
\textbf{(b)}
Spring compression  depends on cell position. 
In equilibrium compression is lowest for the lead cell and highest for the mother cell.
Higher equilibrium spring compressions near the back of the trap reflect the higher mechanical inhibition detected by cells close to the mother cell.
Equilibrium spring compression is a quadratic function of position in the trap (see Appendix).
\textbf{(d)}
Expansion speed depends on cell position.
Although all four cells eventually reach the same steady-state expansion speed, cells near the back of the trap take longer to do so.
Rest length expansion rate was set to $\dot{R} = \SI{0.1}{\micro\meter\per\minute}$ and initial cell length was $\SI{2}{\micro\meter}$.
Spring constant $k$, and damping parameter $\gamma$ were set to $1.0$ as in Figure ~\ref{fig:IsolatedCell}.
}
\label{fig:MM}
\end{figure}

We simulated a mother machine using a single three-walled barrier that allowed cells to grow in a single file.
We initialized cells in the channel by placing them pole-to-pole, with the `mother cell' placed against the back wall (Figure~\ref{fig:MM}(a)). 
As cells grew, they were constrained to move toward the open end of the narrow channel.
Using the model of cell growth described in Section~\ref{s:Framework}, we simulated an array of four cells with constant rest-length extension rate, $\dot{R} = a$, and recorded their resulting spring compressions (Figure~\ref{fig:MM}(b)) and cell-frame expansion speeds (Figure~\ref{fig:MM}(d)).

We see that cell growth rates and spring compressions equilibrate after a transient time determined by the spring constant, resistive damping parameter, and cell position in the mother machine.
This model predicts that the growth rate of the lead cell (the cell closest to the open end of the trap) equilibrates most quickly, and is the least compressed.
This is intuitive, since cells deeper in the trap must overcome the cumulative resistive drag of those nearer the open end.

Analytically, we describe the growing line of cells as a coupled mass-spring system (see Appendix), whose dynamics match the simulations illustrated in Figure~\ref{fig:MM}.
Solving our analytical model shows that steady-state spring compression in a 1D line of cells is a quadratic function of cell position, as Figure~\ref{fig:MM}(b) suggests.

These simulations illustrate cell behavior resulting from competing growth and resistive forces of neighboring cells in a simple geometry.  
Note that steady-state compressions are relatively small in this example.
This is due to the small number of interacting cells, as well as the parameters we selected.
Compressions can grow substantially in larger traps due to increased cell confinement and resulting interaction forces, as we will see in the next section.
Local constraint detection can significantly influence the global dynamics of growing collectives in two-dimensional geometries, as we now demonstrate.

\begin{figure} 
\begin{center}
			\includegraphics[scale=0.96]{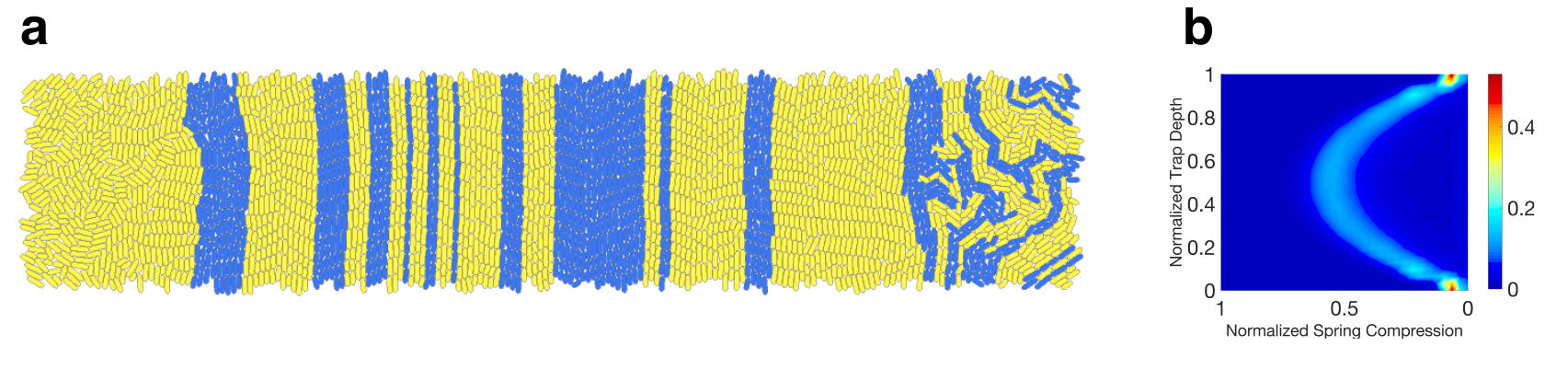}
\end{center}
\caption{
\textbf{Emergent behavior in a long, narrow microfluidic trap.}
We simulated a two-strain consortium growing in a two-dimensional trap open on all sides.
Cells were removed from the simulation once their center of mass crossed a trap boundary.
We initialized the simulation by randomly placing several seed cells from each strain into the trap.
\textbf{(a)}
After the trap was filled, the strains organized themselves into vertically-oriented curvilinear stripes consisting of curvilinear columns of cells.
Each such column functioned as a quasi-mother machine.
\textbf{(b)}
Empirical distribution of normalized cell compression over the length of the simulation.
Each horizontal slice represents the empirical probability density for compression at a given trap depth.
Mean compression is highest in the center of the trap and tapers quadratically, as our theoretical analysis of a mother machines predicts.
The dimensions of the trap were  $\SI{40}{\micro\meter} \times \SI{200}{\micro\meter}$.
All other parameters were set as in Figure~\ref{fig:MM}.
Spring compression is normalized to set the quadratic peak near 0.5.
}
\label{fig:0wall}
\end{figure}

\section{Two-dimensional microfluidic trap geometries: results and predictions}
\label{s:2D}
We next study bacterial assemblies in  two-dimensional geometries.
We start with a two-strain microbial consortium  growing in a long, narrow trap with open sides.
Our model predicts that, after a transient period, strains grow in vertically-oriented, curvilinear stripes perpendicular to the longer edge of the trap.  
Each stripe behaves as a collection of quasi-mother machines. Defects in the stripes form close to the shorter edges of the trap.
While boundary geometry is known to direct the collective orientation of bacterial colonies growing in traps with hard walls~\cite{Cho2007,Volfson:2008}, our prediction of emergent spatiotemporal patterning in open traps is perhaps surprising.
In a final study, we examine how allowing growth rate to depend on spring compression affects the global dynamics of an assembly growing in a trap with three walls. 
Our model predicts that both protein expression and the nematic (angular) ordering of the cells depend on how rest-length extension rate $\dot{R}$ varies with spring compression.

\subsection{Two-strain consortium growing in an open trap}
\label{ss:opentrap}
Agent-based models of cellular growth have provided insights into the spatiotemporal dynamics of collectives~\cite{Jang2012, Boyer2011, Rudge2013}. 
Here, we use our agent-based model to examine the evolving distribution of two strains in a  microfluidic trap open on all sides (see Figure~\ref{fig:0wall}(a)).  
Once a cell reaches the boundary of the trap, we assume that it is rapidly carried away by the flow of the media through a channel surrounding the trap.  
We simulated this by removing such cells from the simulation.  
We initialized the simulation by randomly placing several seed cells of each type into the empty trap.
Cell growth forces were induced by a constant rest-length extension rate, $\dot{R} = a$.

Figure~\ref{fig:0wall}(a) illustrates a typical spatiotemporal pattern that emerges after growth and expansion of the initial seed cells.
Cells organize into vertically-oriented, curvilinear stripes, each composed of a single strain (except for cells near the left and right boundaries, which tend to flow horizontally toward their nearest exits).
Each curvilinear column of cells operates as a quasi-mother machine: Cells at the center of the column act as 'mother cells', while descendants form outer components that flow vertically toward the trap boundary.

Our simulations predict that strain ratio is relatively stable once these stripes emerge.
What determines this stable ratio and the width of the stripes remains unclear, since the transient dynamics that precede this quasi-steady state are complex.
The strain type of the central cell in a given curvilinear cell column determines the strain type of all of the cells in the column.
To predict the stable strain ratio, it is therefore sufficient to predict how the distribution of central cells emerges.  
However, this depends sensitively on the initial distribution of cells, the relative growth rate of the two strains, and other factors ~\cite{Pigolotti2014}.  
Stability of the strain ratio in our simulations emerges from the stability of the quasi-mother machines and their columnar flow, which inhibits cells from lateral motion; notably, only lateral displacement at the \emph{mother cell position} by a different strain can influence the strain ratio non-transiently.

Figure~\ref{fig:0wall}(b) illustrates the empirical distribution of normalized cell compression over the duration of a simulation.
Each horizontal cross-section of this heat map represents the empirical probability density for compression at a given trap depth.
As expected, the empirical compression data is consistent with the behavior of a one-dimensional mother machine.
In particular, mean compression is highest in the center of the trap, and tapers quadratically as one moves to either of the horizontal trap boundaries (we will see that deviations from this quadratic behavior emerge in three-walled traps).
Relatively sharp peaks of the distribution at the long edges of the trap indicate the low-variability of spring compression for cells at the boundary of the columnar flow.

\subsection{Varying the rest-length extension program}
Thus far we have assumed that rest-length extension rate is constant.
We now explore the global implications of allowing rest-length extension rate to vary with spring compression in our model.
This study is motivated by experimental evidence supporting the thesis that mechanical forces shape the dynamics of collectives~\cite{Trepat2009, Sadati2013, Fernandez-Sanchez2015}.
In particular, it has been shown that mechanical forces can become sufficiently large to slow cell growth~\cite{Delarue2016}.
How to best model the impact of such mechanical constraints on cell growth remains unclear.
Here, we therefore  consider a simple model of how cells modulate their target growth rates in response to mechanical forces, and explore the impact of such growth modulation on the emergent properties of the collective.

We introduce a simple growth rate dependence by setting $\dot{R}$ to a constant value for low values of spring compression $C = R - \ell$, while decreasing it linearly to zero after compression crosses a threshold, $T$.
More precisely, we set
\begin{equation*}
\dot{R} (C) =
\begin{cases}
	a, &\text{if } C \leqslant T;
\\
	a(2 - \frac{C}{T}), &\text{if } T < C < 2T;
\\
	0, &\text{if } C \geqslant 2T.
\end{cases}
\end{equation*}
We simulated a three-wall trap geometry, as illustrated in the left column of Figure~\ref{fig:3wall}.
The first row of Figure~\ref{fig:3wall} shows simulation results for a high threshold $T_{h}$ of spring compression, the second for a low threshold $T_{l}$.

The center column (panels (c) and (d)) shows normalized spring compression distributions over the lifespans of the simulations.
The spring compression is normalized such that $T = 0.5$.
As before, a horizontal slice represents the empirical probability density for cell compression at a given trap depth.
Three regimes emerge:
In the bottom section of the trap, the compression profiles are quadratic, suggesting behavior akin to the quasi-mother machine dynamics we examined previously; mean compression levels off beyond the bottom section of the trap before spiking in the back.
The sharply increased spring compression at the back wall emerges from the horizontal alignment tendency of cells in this area.
Cells parallel to the back of the trap have no open trap boundary in their axial growth direction, which results in marked mechanical confinement as evidenced using both thresholds in our simulations.
\subsection*{Implications for protein accumulation}
Spring compression in our model can thus cause cells  within the population to grow at different rates.
This heterogeneity has implications for protein accumulation in growing collectives.
We considered a simple case in which the amount, $x,$ of some protein in each cell obeys the differential equation
\begin{equation*}
\dot{x} = \alpha \ell - \beta x,
\end{equation*}
where $\alpha$ denotes basal production rate and $\beta$ is the rate of chemical degradation. 
When a cell divides, protein is distributed to the daughter cells in proportion to their lengths.
The left column of Figure~\ref{fig:3wall} contains snapshots with cells shaded according to $x / \ell$,  \emph{i.e.} protein per length of cell. As we assumed volume is proportional to length, the shading represented protein concentration within the cells, with brighter cells having a higher concentration of protein.

Protein concentration is highest in the back of the trap,  consistent with the fact that spring compression is highest there.
Significantly more protein accumulation occurs when the threshold $T$ is low (bottom snapshot).
We remark that compression dynamics can be `faster' than protein dynamics in the following sense:
When a cell under significant constraint and expressing a large amount of protein suddenly becomes dislodged (unconstrained), it may  take several generations for protein concentrations in descendant cells to return to levels consistent with equilibrium in unconstrained cells.

\subsection*{Implications for nematic order}
\begin{figure}
\begin{center}
\includegraphics[scale=0.79]{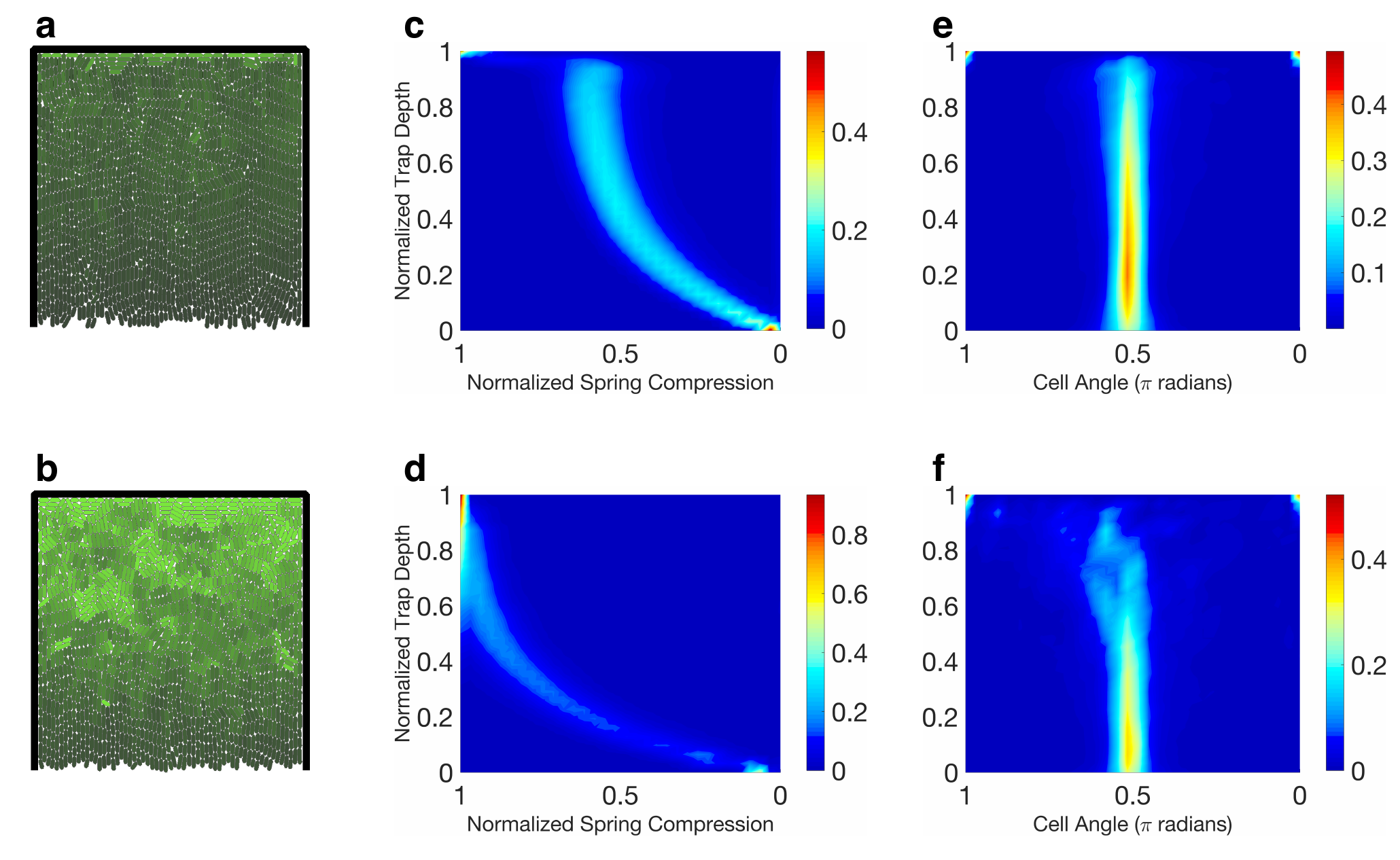}
\end{center}
\caption{
\textbf{Emergent dynamics as a function of rest-length extension rate versus spring compression profile.}
Single-strain collective growing in a three-wall trap.
Rest-length extension rate $\dot{R}$ is constant when spring compression is small, but decreases linearly after compression crosses a threshold.
First row: high threshold; second row: low threshold.
\textbf{(ab)}
Concentration of a constitutively-produced protein (brighter color indicates greater concentration.)
Protein concentration is highest in the back of the trap because typical cell division time is longest there.
Lowering the compression threshold leads to a significant increase in protein accumulation.
\textbf{(cd)}
Distribution of normalized compression over the lifespans of the simulations.
Each horizontal slice represents the empirical probability density function for compression at a given trap depth.
\textbf{(ef)}
As in (cd), but for cell angle instead of compression.
Notice that nematic disorder in the back half of the trap is greater with lower compression threshold.
Trap dimensions: $\SI{65}{\micro\meter} \times \SI{65}{\micro\meter}$.
}
\label{fig:3wall}
\end{figure}

We finish by examining how nematic order is affected by altering the rest length extension rate vs. spring compression profile.
The right column of Figure~\ref{fig:3wall} shows cell angle distributions over the lifespans of the simulations.
An angle of $\pi / 2$ corresponds to a vertically oriented cell.
Each horizontal slice in the figure represents the empirical probability density function for cell angle at the given trap depth.
When the threshold $T$ is high, as in Figure~\ref{fig:3wall}(e), cells show strong vertical alignment throughout the trap.
We observe significantly more nematic disorder with a lower threshold (Figure~\ref{fig:3wall}(f)).

Boyer et al.~\cite{Boyer2011} have shown that nematic disorder in three-wall trap geometries can be caused by a buckling instability.
Under the assumption that cells in the back of the trap both slow their growth and are smaller due to nutrient depletion, they further show that nematic disorder will be more prevalent there since small cells are more likely to buckle (Figure~5 of~\cite{Boyer2011}).
By reducing $T$ in our simulations, we observe that reduction of cell growth rate alone leads to strong nematic disorder in the back of a three-walled trap geometry.
Consequently, we have recapitulated the Boyer result.  However, in our case the mechanisms are different:  Nematic disorder emerges solely from slowing cell growth rate, which follows directly from detection and response to mechanical interactions, and not from postulated nutrient depletion.

\section{Discussion}
The growth of cells, both in natural environments and experimental conditions, is modulated by a number of factors.  
These include mutations, nutrient depletion, extracellular forces, and other environmental signals. 
Cells actively respond to mechanical forces, which implies they are capable of sensing and transducing these signals to a biological response ~\cite{Hatami-Marbini2011}.
Here, we have described a simple model of how bacteria effect changes in their growth in response to mechanical interactions.
We have shown that such changes can impact the spatiotemporal dynamics of bacterial collectives growing in microfluidic traps. 

However, our model is certainly an oversimplification.  
We did not attempt to describe the other factors that modulate cell growth and can lead to emergent dynamical phenomena. 
For instance, assume that the growth rates of two co-repressing strains in a consortium depend on their transcriptional states, so that the strain that has the higher level of expression grows more slowly. 
This type of interaction between cell growth, strain competition, and protein expression can lead to relaxation oscillations in both transcriptional and growth rates~\cite{Sadeghpour:2017}.  
We expect that a variety of mechanisms that affect growth rate of single cells directly or indirectly can lead to emergent phenomena at the level of the bacterial population.

Our agent-based model stands in contrast to most previously developed models: We allow cells to follow first-order dynamics rather than assuming cells achieve their growth rate in each time step. 
Thus cells effectively monitor their environment and respond to mechanical interactions by modulating growth, and, potentially, other aspects of their interior dynamics. 
It is unclear whether a cell that is prevented from growing stores this potential. 
However, mechanical interactions certainly impact cell growth even when nutrient supply is adequate.  
This is confirmed by experiments performed in osmotic shock, where not only do cells no longer grow, they also return to the cell length they would have achieved, had shock not occurred~\cite{Huang:2008}.  
Thus, growth is "stored," and this also supports the use of increased spring compression in our model under constraint. 

Although our model is an oversimplification, it shows that  mechanical interactions can play an important role in the organization and dynamics of growing bacterial collectives.  
We have described a flexible platform for understanding these effects.
But much work remains: The predictions of these models, such as the organization of colonies in microfluidic traps, and the impact of crowding on gene expression will need to be validated experimentally.
A deeper understanding of the emergence of order and disorder in these bacterial populations will require the development of effective continuum models of collective cell dynamics~\cite{Volfson:2008}.  
Agent-based models of the type we describe can serve as a starting point for these further developments.


\section{APPENDIX}
\subsection			{NON-INERTIAL DYNAMICS ASSUMPTION}
\label{sm:noninertial}
The non-inertial dynamics assumption is satisfied in a regime defined by the value of a a fast-scale time constant $\xi$, which we define with respect to the \emph{inertial} dynamics equations of motion for an isolated cell in our model.
We begin with the assumption that the expansion force on a cell is constant. In our model, this translates to a fixed compression $R - \ell$ of our expansion spring.  
The validity of this assumption for our simulations is validated by the scale difference between $\xi$ and the discretization time step $dt$ (during which we assume expansion force is constant).
We will see that $dt$ is much larger than $\xi$. 

Referring to Eq.~\eqref{eqn:inertial} in section ~\ref{s:Framework} , we set $F^{exp} := 2(R - \ell)$ as the constant expansion force. 
Assuming the mass, $m$, is constant, the inertial equation of motion for an expanding cell in our model is then:
%
\begin{equation}
	\ddot{\ell} = {F^{exp}\over m} - {\gamma\over m} \dot{\ell}.
\end{equation}
%
We define $\xi := {m\over \gamma}$ as our fast-scale time constant.  Solving this equation with initial velocity $\dot{\ell}(0)$ at time $t=0,$ we obtain the expansion velocity solution,
%
\begin{equation} \label{eqnSM:xdot}
	\dot{\ell}(t) 
	= (1 - e^{-{t\over \xi}}) \cdot \frac{F^{exp}}{\gamma} \enspace 
	+ \enspace     e^{-{t\over \xi}} \cdot \dot{\ell}(0).
\end{equation}
%
Thus, for times $t$ under which our constant force assumption holds,  the cell expansion velocity is a convex combination of its \emph{terminal velocity} and initial condition.  We can now compute an explicit equation for the acceleration of the cell by taking the time derivative of (\ref{eqnSM:xdot}):
%
\begin{equation}
	\ddot{\ell}(t) 
	= {d\over dt} \dot{\ell}(t) 
	= e^{-{t\over \xi}} (\frac{F^{exp}}{m} - \frac{\gamma}{m} \dot{\ell}(0))
\end{equation}
%
Thus, at $t = 0$, the acceleration is \emph{inertial} and decays exponentially.  
From equation (\ref{eqnSM:xdot}), we thus see that \emph{non-inertial dynamics} holds to the extent that $F^{exp}$ can be assumed constant over a time interval $t$ of interest, such that (conservatively) $t \geq 10\xi$ (the exponential decays to $< 10^{-4}$ in this time).
If we take the mass of a cell as $m_{cell} = 10^{-15}$ \texttt{kg} and a fluid damping parameter $\gamma = 10^{-8}$ \texttt{kg/sec}, we have $\xi = 10^{-7}$ \texttt{sec}, or $0.1 \mu$ \texttt{sec}.
Our computer simulations use a time discretization on the order of $dt = 0.001$ \texttt{min} $= 0.06$ \texttt{sec}. 
We then have $dt/\xi > 10^5$.  

Thus, during a simulation time interval $dt$ (under which we assume spring rest-length and cell mass are constant), our non-inertial dynamics assumption holds.  
Indeed, assuming the given cell mass and fluid damping values, non-inertial dynamics holds whenever system forces and masses can be assumed constant over time intervals of $\mu$\texttt{sec} or greater.
\subsection			{TIME DISCRETIZATION REQUIREMENTS}
\label{ss:dt}
Under the non-inertial dynamics assumption (see section ~\ref{sm:noninertial}) , expansion velocity is proportional to expansion force.  
In order to prevent overshoot of the expansion velocity for an isolated cell in our simulations, we must observe an upper bound for our discretization time step $dt$.  
To see this, we require that $\dot{\ell} < a$.  
That is, the achieved expansion speed of a cell (starting from rest) should be less than the cell growth rate $a$.  
In the RHS of equation (\ref{eqn:noninertial}), we set $t = dt$ to perform a forward Euler integration of the rest length (thus $= adt$).
We set $\ell(0) = 0$ and conclude:
%
\begin{equation} \label{eqnSM:dt}
	\dot{\ell} < a 
	\implies  \frac{2k}{\gamma} a dt < a  
	\implies dt < \frac{\gamma}{2k} = \frac{\tau}{2}
\end{equation}
%
Thus, $dt<\frac{\tau}{2}$ is a necessary condition in our discretization to prevent expansion speed overshoot from rest.
Importantly, this directly links the lower range of $\gamma$ (for fixed $k$) to computation time: increased computation time is the result of a smaller $dt$, which is required by a smaller $\gamma$.
Thus, simulations that explore smaller values of $\tau$ (equivalently, smaller values of $\gamma$ for fixed $k$) will engender  higher computational cost under the model described in this paper.
However, a more sophisticated, nonlinear control scheme to regulate expansion velocity could be implemented to mitigate this restriction.  
Here we retained a simple open-loop growth algorithm  to validate agreement between theory and our simulation environment, leaving the development of more advanced control algorithms for future work. 

\subsection			{COUPLED MASS-SPRING MATRIX EQUATIONS}
We now derive the equations of motion for a 1D line of bacterial cells using our model's mass-spring system.  
In this derivation, we also include the possibility of \emph{spring damping}, which is a cell-frame dashpot damping added to the expansion spring.  We analyze the impact of this damping on the resulting dynamics.

\subsubsection		{3-CELL MOTHER MACHINE}
We assume a 1D line of 3 cells in a mother machine configuration (see section ~\ref{s:1Dmm}) where cells are in contact pole-to-pole and are constrained to motion in the axial direction only.
Since each cell is composed of two axially-independent halves, the mother machine configuration will identify positions of the contacting cell halves of adjacent cells.
The mother cell's trap-wall half will not move in this configuration, thus the equations of motion are determined for the identified positions of each successive cell-cell contact ($i=1,2$) and lastly for the free-end cell half ($i=3$), where $i$ is the index number for the equations given below.

We assume a spring constant $k$, fluid damping parameter $\gamma_f$, and spring damping parameter $\gamma_s$.  
The matrix-vector equations for an example 3-cell mother-machine system are generated by a stiffness matrix $\bold{K}$ and damping matrix $\bold{\Gamma}$, which are second-difference matrices that follow from force-balance analysis ~\cite{strang2007} of the 1D line of masses and springs that represent a back-to-back line of cells in a mother machine using our model.  
We find
%
\begin{equation}
	 \bold{K} = 
  \begin{bmatrix}
    \ \ 2 & -1 & \ \ 0 \\
   -1 & \ \ 2 & -1 \\
    \ \ 0 & -1 & \ \ 1  \end{bmatrix}
    , \qquad and \qquad
     \bold{\Gamma} =
  \begin{bmatrix}
    2\gamma_s + \gamma_f & -\gamma_s & 0 \\
    -\gamma_s & 2\gamma_s + \gamma_f & -\gamma_s \\
    0 & -\gamma_s & \gamma_s + \gamma_f  \end{bmatrix} 
    .
\end{equation}
%
The equations of motion for the coupled system from Newton's 2nd Law are:
%
\begin{equation} \label{eqn:ddotX}
	 m\bold{\ddot x} = -k\bold{K x} - \bold{\Gamma \dot x} 
	 + k \left [ \begin{array}{c} 0\\  0\\ 1\end{array} \right ] a t
\end{equation}
%
where $a$ is the cell growth rate.  
Cell 1 is the mother cell and cell 3 the open-end cell in the mother machine. 
Internal force cancellation of adjacent cell halves results in the RHS of the above equation having a forcing term only for the outermost cell half of the open-end cell.
Expansion forces are then realized through coupling in the stiffness matrix $\bold{K}$.
Assuming non-inertial dynamics  (see section ~\ref{sm:noninertial}) and that $\bold{\Gamma}$ is invertible, (\ref{eqn:ddotX}) becomes:
%
\begin{equation} \label{eqn:dotX}
	  \bold {\dot x} = 
	  -k\bold{\Gamma^{-1}} \bold{K x} 
	 + k \bold{\Gamma^{-1}} \left [ \begin{array}{c} 0\\  0\\ 1\end{array} \right ] a t
\end{equation}
%
Now, assuming the matrix product $\bold{\Gamma^{-1}} \bold{K}$ is diagonalizable with eigenvector matrix $\bold{Q}$, we have the equivalent system of equations in the eigen-basis (using the vector variable $\bold{y}$ in this basis):
%
\begin{equation} \label{eqn:ydot1}
	  \bold{\dot y} = -k\bold{Q^{-1}(\Gamma^{-1}} \bold{K}) \bold{Q y} 
	 + k \bold{\bold{Q^{-1}} \Gamma^{-1}} \left [ \begin{array}{c} 0\\  0\\ 1\end{array} \right ] a t
\end{equation}
%
If we set 
$\bold{b} :=  k\bold{\bold{Q^{-1}} \Gamma^{-1}} \left [ \begin{array}{c} 0\\  0\\ 1\end{array} \right ] $, 
 with diagonal eigenvalue matrix $\bold{D}$, the diagonalized matrix-vector equation becomes:
%
\begin{equation} \label{ydot2}
	  \bold{\dot y} = -k\bold{D y} 
	 + \bold{b} a t
\end{equation}
%
The solution to the diagonalized system now follows as for the single-cell case given by Eq.(\ref{eqn:ell}) in section (\ref{s:Framework}).  
For $i \in \{1,2,3\}$, we set $\tau _i := \frac{1}{k\bold{D}_{ii}}$, and have:
%
\begin{equation} \label{eqn:ifd}
	 \partial_t (e^{\frac{t}{\tau _i}} y_i) = e^{\frac{t}{\tau}} b_i a t
\end{equation}
%
Assuming each $y_i(0) = 0$, we then have the diagonalized solutions:
%
\begin{equation} \label{eqn:yi}
	  y_i = \tau _i b_i a ( t -\tau _i + \tau _i e^{- \frac{t}{\tau _i}})
\end{equation}
%
\begin{equation} \label{eqn:yidot}
	  \dot{y_i} = \tau _i b_i a ( 1- e^{- \frac{t}{\tau _i}})
\end{equation}
%
We then convert the solution back to the standard basis using $\bold{x} = \bold{Qy}$ and  $\bold{\dot x} = \bold{Q \dot y}$.
We thus have that the motion of each cell in the mother machine is a linear combination of eigen-modes of the matrix product $\Gamma^{-1} \bold{K}$.
We now explore the effects of the spring damping on the equations of motion.

\subsubsection		{NO SPRING DAMPING}
With no spring damping, the $\bold{\Gamma}$ matrix is diagonal and we can replace it with a scalar parameter $\gamma$.   
$\bold{Q, D}$ are then the eigenvector, eigenvalue matrices of $\bold{K}$, and we set:
\begin{equation} \label{eqn:ifd}
 \tau _i := \frac{\gamma}{k D_{ii}}, \qquad b_i :=  \frac{k\bold{Q}^{-1}_{i3}}{\gamma}
\end{equation}
%
The solution is then given by (\ref{eqn:yi} - \ref{eqn:yidot}).
We find that the steady-state solutions to spring compression follow a quadratic profile vs. cell position in the mother machine.
This is readily derived without the matrix equations by analyzing the the force balance necessary to achieve linear growth in cell-end speeds towards the open end of the mother machine.
If we assume each cell expands (in the cell's frame of reference) at a constant speed $v$, then each successive cell-end will move (in the laboratory frame of reference) at $i \cdot v$, where $i \in 1..N$ and $N$ is the number of cells in the mother machine, and $i=1$ is the mother cell.
Since the end-cell half of cell $N$ is independent and we are under non-inertial dynamics, this end-cell half must apply a force of $\gamma N v$ to achieve speed $N v$ in the laboratory frame.
Each cell expands with symmetric force, thus cell end $(N-1)$, which moves at $(N-1) v$ must have (by algebraic addition of forces from adjacent cell halves):
$\gamma C_{N-1} v  - \gamma N v = \gamma (N-1) v$,
where $C_{N-1}$ is the unknown scale factor for the pentutimate cell.  
Clearly, $C_{N-1} = N + (N-1)$.

\begin{figure}[ht]	
\centering
			\includegraphics[scale=.35]{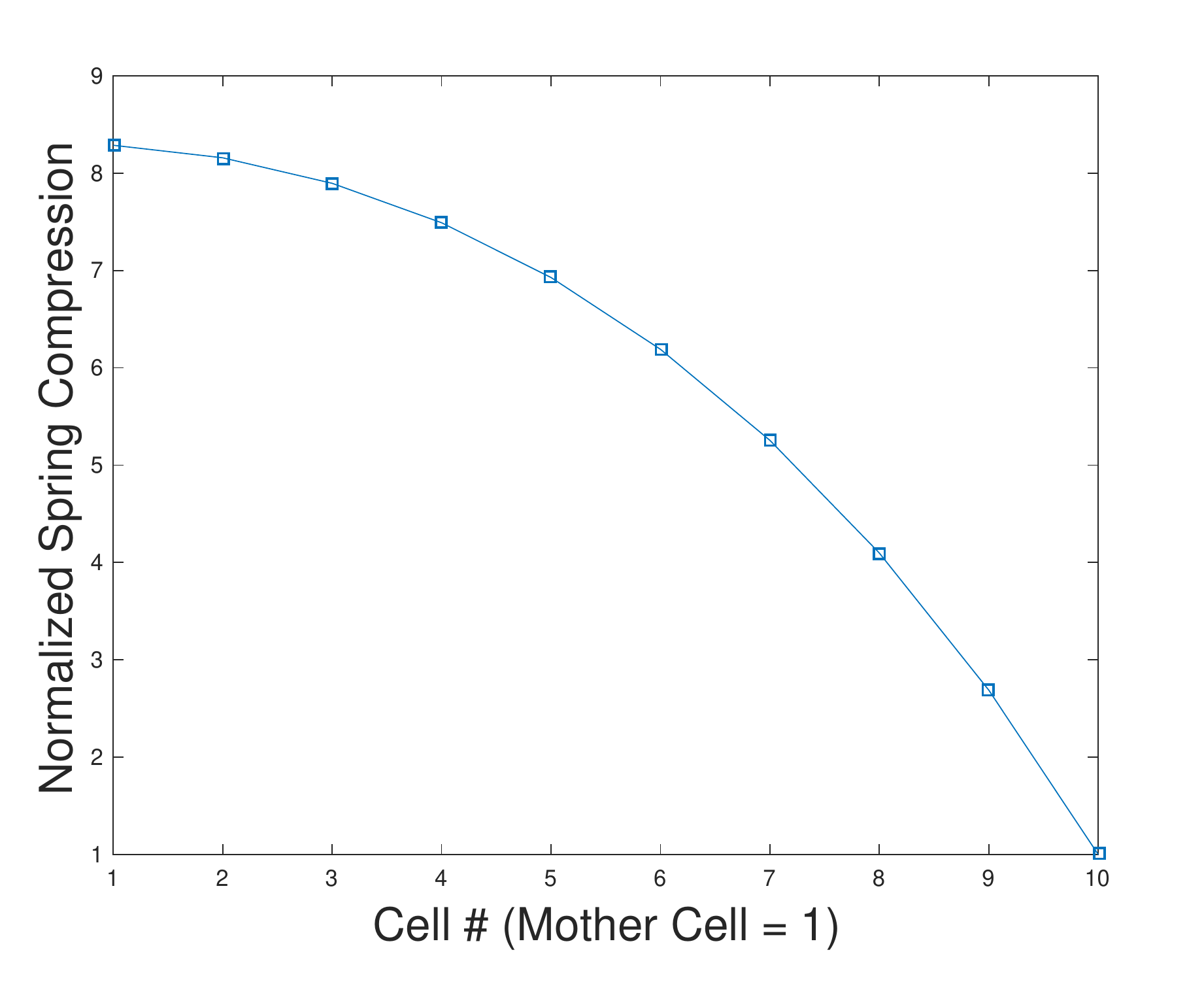}
\caption{
Normalized steady-state cell compression for $N=10$ cells in a mother-machine.  
The quadratic profile is predicted by the analytical solution.
}.
\label{fig: MMquad}
\end{figure}

Continuing in this manner towards the mother cell, we see that the successive cell force differences lead to a quadratic expression for cell compression vs. cell position.
An example plot of the steady-state cell compression for $N=10$ cells is shown in Fig.(\ref{fig: MMquad}), where the quadratic profile is evident.

\subsubsection			{LARGE SPRING DAMPING}
With spring damping much larger than fluid damping, we then have for the damping matrix:
\begin{equation}
	 \bold{\Gamma} = \gamma _s
  \begin{bmatrix}
    \ \ 2 & -1 & \ \ 0 \\
   -1 & \ \ 2 & -1 \\
    \ \ 0 & -1 & \ \ 1  \end{bmatrix}, \qquad 
\end{equation}
We note in this case that $\bold{\Gamma} = \gamma _s \bold{K}$.
Importantly, increasing spring damping relative to fluid damping leads to \emph{uniform dynamics} of all cells in the mother machine.
However, to maintain responsiveness, the spring constant $k$ must scale with spring damping.
For example, in the isolated cell case, spring damping and fluid damping reference frames are the same, and the damping parameters add such that $\gamma = \gamma_f + \gamma_s$ to define $\tau$ in Eq. (\ref{eqn:ell}).
Thus, to maintain the same first-order dynamics, $k$ must scale with the resulting additive $\gamma$ such that $\tau = \frac{\gamma}{k}$ remains constant.
We find that, in the limit of $k,\gamma_s \to \infty$, while $\frac{\gamma_s}{k} = constant$ we recover the EOR model described in section ~\ref{s:Framework}.

\subsection			{CHIPMUNK 2D SIMULATION ENVIRONMENT}
We use the open-source physics engine \texttt{Chipmunk 2D} (See \cite{cp}) to define cell objects and  traps to simulate interactions and dynamics of cell consortia.  The use of this engine by \texttt{gro} (See \cite{Jang2012}) was the original inspiration for its use in our model.
We detail in this section our simulation loop algorithm and the relevant components from \texttt{Chipmunk 2D}.  

A simulation step consists of the following. 
The 2D physics engine is assumed to have just completed a time-step. 
An un-ordered list structure of cell objects is then traversed to determine if a cell should divide or be removed from the simulation (sub-routines would either add a new daughter cell or remove the cell from the list, respectively). 
Each remaining cell's physics model is then updated as follows:
\begin{enumerate}
\item  
The current cell length $\ell$ is computed by subtracting the positions of the cell ends, which are obtained by querying the respective components from the 2D physics engine.  The current spring compression is then computed by subtracting the cell length from the spring rest length.
\item  
As a function of the current spring compression, a growth rate is selected for the following time step (the growth rate may also be constant, i.e., independent of compression, or in general, it may be set algorithmically by the user).  
The growth rate is then asserted in the discrete-time simulation by an increase of the spring rest-length, with increment $dR := a \cdot dt$, where $a$ is the current growth rate and $dt$ is the discrete time step.  
Thus, $R \gets  R + dR$.
\item  
The cell expansion force is computed as $F^{exp} = k(R - \ell)$ and this is set for each cell half independently.  
\item  
The 2D physics engine is stepped.  This consists of 3 principal parts within the \texttt{Chipmunk 2D} software:
\begin{enumerate}
\item
The current timestep velocity $\bold{v}_i$ is forward-integrated to determine new positions for all objects $i$ in the space.
Namely, each cell end is extended by $\bold{v}_i \cdot dt$, where $\bold{v}_i$ is computed at the end of the previous physics engine time step (or otherwise altered by the user in steps 1-3 above).  
In general, cells will \emph{not overlap} each other as the result of a position integration.  
Rather, objects in contact will move together with velocities that were resolved in the previous time step (in part C below) by the physics engine via collision dynamics.
\item
The force programmed in item (3) above is used to determine new interaction velocities for all objects in the space. 
The cell halves' velocities in the non-inertial regime are computed directly by $\hat{\bold{v}} = \frac{\bold{F}^{exp}}{\gamma}$. (The previous velocity of the objects is set to zero in the non-inertial regime). 
In general this velocity $\hat{\bold{v}}$ will not actually be achieved by the cell halves.
The impulse solver in part C will adjust velocities and positions based on collision dynamics of objects in the space.
\item
The 2D physics engine's impulse solver iterates over the space to resolve competing object velocities when two objects are in direct contact.  
The impulse solver adds impulses to each object and the resulting actual velocities of cell halves are computed and will be used in the following timestep's position integration.
\end{enumerate}
\end{enumerate}

\subsection		{RATCHET ALGORITHM FOR CELL BACK-FILLING}
To ensure that axial compression is accounted for in our model, we employ an algorithm to back-fill contact area to each cell half, such that contraction of a cell is limited to a compression gap and ratchet step, which we now detail.  
Each cell half is constructed as in Fig. ~\ref{fig:IsolatedCell} with a rectangular center and attached ``frontside pole'' that defines the frontal contact area of a cell.
In addition, however (and not shown in Fig. ~\ref{fig:IsolatedCell}), there is a ``backside pole'' that is attached to a ratchet-extended rectangular area, which is designed to keep the backside pole just inside the frontside pole of the other half.
Usually, this backside pole contact surface is \emph{transparent} to collision dynamics of a cell, since it lies inside the outer contact hull of the cell.
However, in case a cell becomes subject to constricting axial forces (from other cells or trap walls) larger than cell expansion forces, the two halves will contract, but only until the backside poles of each half align with the frontside poles of the other, at which time the cell acts as a rigid body not subject to further compression.

As a cell expands, this backside contact area must be extended to limit the amount of compression before the poles are aligned from the two halves (thus forming the rigid body).
We employ a ratchet algorithm to achieve this extension, such that the backside pole is extended once the cell length passes a ratchet step $r_s$.  The algorithm is summarized as follows:
\begin{algorithm}[ht]
\SetAlgoLined
	\KwData{
		current cell length $\ell$; 
		current ratchet count $n$; 
		ratchet step $r_s$; 
		ratchet gap $r_g$; 
	}
	\KwResult{
		on next ratchet: back-filled cell contact area to $\ell - r_g$;
	}
	Initialize each cell on birth with backside pole $r_g$ away from frontside pole of other half and set $n=0$\;
	For every cell in every time step:
	
	\eIf{$(\ell > (n+1)r_s + r_g)$}{
		back-fill cell pole contact area to $\ell - r_g$ for each cell half\;
		$n \gets n+1$\;
	}{
		continue\;
	}

\end{algorithm}

\subsection		{TABLE OF PARAMETER VALUES}
~\newline
\begin{center}
	 \begin{tabular}{|c | c | c | c|} 
	 \hline
	 PARAMETER & SIMULATION VALUE & SCALE & PHYSICAL VALUE \\ 
	 \hline
	 $dt$ &  $\SI{0.001}{min}$ & 1 &  $\SI{0.06}{sec}$ \\ 
	 \hline
	 m & $\SI{1e-10}{}$  &  $\SI{1e-5}{kg}$ & $\SI{1e-15}{kg}$ \\
	 \hline
	 $\gamma$ & $\SI{60}{min^{-1}}$ & $\SI{1e-5}{kg}$ & $\SI{1e-5}{kg.sec^{-1}}$ \\
	 \hline
	 k &  $\SI{3600}{min^{-2}}$ & $\SI{1e-5}{kg}$ & $\SI{1e-5}{kg.sec^{-2}}$ \\
	 \hline
	 $\xi := m \gamma^{-1}$ & $\SI{1.7e-12}{min}$ & 1 & $\SI{1e-10}{sec}$\\ 
	 \hline
	 $\tau := \gamma k^{-1}$ & $\SI{0.017}{min}$ & 1 & $\SI{1}{sec}$ \\ 
	 \hline
	\end{tabular}
\end{center}
~\newline

Simulation values are computed by dividing the physical value by the scale for each parameter and converting units appropriately.
We used mass $m$ of a bacterial cell as given in ~\cite{bionum}, and dimensionless mass units in our simulations (the scale value for mass is chosen to normalize $k$ and $\gamma$ to $1.0$ in SI units).
Our estimate for $k$ derives from the data given in ~\cite{Huang:2008} (Supporting Information).
We chose $\gamma$ such that $\tau = \SI{1}{sec}$. 
In Fig.~\ref{fig:IsolatedCell}, panels (b),(c), we also used a value of $10 \gamma$ for comparison of the cell dynamics.
Both $k$ and $\gamma$ simulation values use dimensionless mass units.
Two time constants are shown for reference: $\xi$ defines a scale for non-inertial dynamics as in \ref{sm:noninertial}, and $\tau$ defines the first-order growth dynamics of our cell model, as derived in \ref{ss:eom}.
We note that $\gamma$ overestimates a physical value, but it is chosen as a convenient value for computational purposes (see \ref{ss:dt}).
  


{\bf Acknowledgments:} 
This work was supported by the National Institutes of Health, through grant R01GM117138 (MRB, KJ, WO) and the joint NSF-National Institute of General Medical Sciences Mathematical Biology Program grant R01GM104974 (MRB, KJ, WO); NSF grants DMS-1413437 (WO), MCB-1616755 (OI), and MCB-1411780 (OI); and the Welch Foundation grant C-1729 (MRB).

The authors acknowledge the use of the Opuntia Cluster and the support from the Center of Advanced Computing and Data Systems at the University of Houston.


\bibliographystyle{siam}	
 \bibliography{./bib/jmech,./bib/jconsortia,./bib/jsim,./bib/jjamming,./bib/jsoft,./bib/jgrowth2,./bib/jcolony2,./bib/basket,./bib/jvar}

\end{document}